\documentclass[floatfix,twocolumn,showpacs,prd,amssymb,amsmath]{revtex4}

\usepackage{graphicx}
\usepackage{amsbsy}
\usepackage{aas_macros}
\usepackage{bm}% bold math

\def\be{\begin{equation}}
\def\ee{\end{equation}}
\def\bea{\begin{eqnarray}}
\def\eea{\end{eqnarray}}

\def\bi#1{\mathbf{#1}}
\def\bs#1{\boldsymbol{#1}}
\def\D{\bi{D}}
\def\beps{\bs{w}}
\def\C{\bi{C}}
\def\bS{\bi{S}}
\def\Z{\bi{Z}}
\def\A{\bi{A}}
\def\B{\bi{B}}
\def\Sg{\bs{\Sigma}}
\def\Sgi{\Sg^{-1}}
\def\t{^{\dag}}
\def\like{\mathcal{L}}

\def\a{\mathbf{a}}
\def\b{\mathbf{b}}
\def\c{\mathbf{c}}
\def\E{\mathbf{E}}

\def\f{\mathbf{f}}
\def\g{\mathbf{g}}
\def\h{\mathbf{h}}
\def\F{\mathbf{F}}
\def\p{\mathbf{p}}
\def\q{\mathbf{q}}

\def\Sgc{\Sg^\C}
\def\SgB{\Sg^\B}
\def\in{^{-1}}

\def\WMAP{{WMAP}}
\def\citeN#1{\cite{#1}}

\begin{document}
\newlength{\figwidth}
\setlength{\figwidth}{\columnwidth}
 
\title{Sunyaev-Zeldovich effect in \WMAP\ and its effect on cosmological 
parameters}

\author{Kevin M. Huffenberger} \email{khuffenb@princeton.edu}
\author{Uro\v s Seljak}
\author{Alexey Makarov}
\affiliation{Department of Physics, Jadwin Hall\\ Princeton University, 
Princeton, NJ 08544}

\date{\today}

\begin{abstract}
We use multi-frequency information in first year \WMAP\ data 
to search for the Sunyaev-Zeldovich 
(SZ) effect. WMAP has sufficiently broad frequency coverage to constrain SZ 
without the addition of higher frequency data: 
the SZ power spectrum amplitude is expected to increase 50\% from 
W to Q frequency band. This, in combination with the low noise in \WMAP, allows
us to strongly constrain the SZ contribution. We derive an optimal frequency 
combination of \WMAP\ cross-spectra to extract SZ in the presence of noise,
CMB, and radio point sources, which are marginalized over. 
We find that the SZ contribution is 
less than 2\% (95\% c.l.) at the first acoustic peak in W band. 
Under the assumption that the removed radio point sources 
are not correlated with SZ this limit implies $\sigma_8<1.07$ at 95\% c.l.
We investigate the effect on the cosmological 
parameters of allowing an SZ component. We run Monte Carlo Markov Chains with 
and without an SZ component 
and find that the addition of SZ does not affect any of the cosmological 
conclusions.
We conclude that SZ does not contaminate the \WMAP\ CMB or change cosmological
parameters, refuting the recent claims that they may be corrupted.
%, with the largest effect being a 0.5$\sigma$ increase in $\Omega_bh^2$.  
\end{abstract}

\pacs{98.65., 98.65.Dx, 98.65.Hb, 98.70.Vc, 98.80., 98.80.Es}

\maketitle

\section{Introduction}
{\em Wilkinson Microwave Anisotropy Probe} \cite{2003ApJS..148....1B} (\WMAP) 
observations of the Cosmic Microwave Background (CMB) have ushered in a new 
era of high precision observational cosmology. Such a tremendous 
increase in data quality requires a corresponding increase in the care 
that goes into the data analysis and interpretation. One of the lingering 
concerns surrounding the analysis is the residual effect from additional 
sources of anisotropies. A very prominent candidate among these 
is the Sunyaev-Zeldovich effect (SZ). 
Electrons in hot gas scatter photons and 
distort the blackbody spectrum of the CMB.  
  Galaxy clusters, where gas is the hottest,
 contribute the bulk of the effect, shock-heating the gas in 
their potential wells.  
Radio and microwave band observations pointed at known clusters 
routinely yield SZ detections, so this effect is now well established
observationally.  
The scattering preferentially raises the energy of the CMB photons, 
but the number of scatterings is low, so the process never achieves thermal 
equilibrium.  Therefore SZ appears as a CMB temperature decrement at low 
frequencies and as an increment at high frequencies.  
This frequency dependence is well known: it is 
constant in the Rayleigh-Jeans (RJ) regime of the blackbody spectrum, and is 
universal (independent of gas temperature or density) 
for non-relativistic electrons.

In the channels of \WMAP, SZ is a temperature decrement.  Table 
\ref{tab:wmapbands} provides \WMAP's frequency bands, and the SZ frequency 
dependence in those bands.  The K and Ka bands are the most heavily polluted by
 galactic contamination, so the best prospects for identifying SZ in WMAP are 
in the differencing assemblies of the upper three bands: two assemblies in Q, 
two in V, and four in W.
While it is usually argued that SZ is indistinguishable from the CMB in the  
\WMAP\ channels this is actually not so: in CMB temperature units 
the SZ power spectrum increases by 
50\% from W to Q channel and in W it is only 61\% of the RJ power. 
We will show that this suffices to place strong constraints on SZ using 
\WMAP\ data alone, a consequence of the remarkably low noise in \WMAP\ data. 
\begin{table}
\begin{center}
\begin{ruledtabular}
\begin{tabular}{lrrr} % \\
  Band& $\nu$ (GHz) & ${\Delta T^\text{SZ}/(T^\text{CMB} y)}$ & 
$C_l/C_l^\text{RJ}$\\ \hline % & & & \\
K    &   23             &   $-1.97$ &   $0.972$  \\ %  & & & \\
Ka   &   33             &   $-1.94$ &   $0.945$  \\ %  & & & \\
Q   &    41             &   $-1.91$ &   $0.915$  \\ %  & & & \\
V   &   61              &   $-1.81$ &   $0.820$  \\ %  & & & \\
W   &   94              &   $-1.56$ &   $0.611$  \\ %  & & & \\
\end{tabular}
\end{ruledtabular}
\end{center}
\caption{SZ contribution in the bands of \WMAP.  We note the band name, 
frequency, temperature perturbation relative to the comptonization parameter 
$y$, which does not depend on frequency, and power spectrum relative to the 
power in the Rayleigh-Jeans regime.  In the \WMAP\ bands, SZ is a temperature 
decrement.  In the Rayleigh-Jeans regime, ${\Delta T^\text{SZ}/(T^\text{CMB} 
y)}=-2$. }\label{tab:wmapbands}
\end{table}

In the literature, several groups have attempted to identify SZ in \WMAP\ data 
using cross-correlations with other tracers of large scale structure (LSS). In 
\citeN{2003ApJS..148...97B} they 
build an SZ template from the XBACs catalog of x-ray clusters 
\cite{1996MNRAS.281..799E}, and fit for the amplitude of SZ, arguing for a 
$2.5\sigma$ detection.
In \citeN{2003astro.ph..8260A} they compute the cross-power spectra  of the 
2MASS extended source catalog \cite{2000AJ....119.2498J} with \WMAP's Q, V, and
 W bands, then fit for the amplitude of the SZ signal, arguing for a 
3.1--3.7$\sigma$ detection of SZ, depending on their mask.
In \citeN{2003MNRAS.346..940D} they 
compute the cross-power spectrum of the ROSAT diffuse x-ray background maps 
\cite{1997ApJ...485..125S} and  a \WMAP\ (Q-W) difference map, finding no 
detection.
In \citeN{2003ApJ...597L..89F} they compute the cross-correlation function of 
SDSS DR1 \cite{2003AJ....126.2081A} galaxy survey data with the V band and fit 
for the amplitude of SZ, finding a 2.7$\sigma$ detection. In 
\citeN{2004MNRAS.347..403H} they 
build templates for SZ from several optical and x-ray cluster surveys, then fit
 for the amplitude of these templates using maps from the W band and 
\citeN{2003PhRvD..68l3523T}, finding
no detection for optical clusters and 2--5$\sigma$ detections for x-ray 
clusters.

Finally, in \citeN{2004MNRAS.347L..67M} they compute the mean temperature in 
concentric rings about APM \cite{1990MNRAS.243..692M} and ACO 
\cite{1989ApJS...70....1A} groups and clusters, noting a decrement they 
attribute to SZ.  They interpret this decrement to extend to large scales, 
$\sim 1$ degree, although the covariance of their bins is unclear.  Redshift 
$z<0.2$ clusters dominate their sample.  Their most extreme model assumes that 
extended SZ emission is representative of the temperature and spatial 
clustering of gas to $z=0.5$.  In this model the SZ power is 30\% of the first 
acoustic peak of the CMB.  \
Thus they
conclude the integrity of the \WMAP\ cosmological parameter fits may be 
compromised. 

In this paper we seek to place limits on SZ from \WMAP\ data alone, thus 
avoiding any uncertainties in connecting the results of \WMAP-LSS 
cross-correlation 
to those of \WMAP\ auto-correlation.  The cross-correlation methods 
used before require a model for relating SZ temperature perturbations to some 
proxies for SZ, which are nonlinear structures such as clusters, and this 
connection 
can be quite uncertain.  Our method is less model dependent.  
The downside is that \WMAP-only methods sacrifice signal to noise because they 
do not focus on matter concentrations, where SZ should be strongest, so our 
method is not optimized for obtaining an SZ detection.
However, our main goal is to investigate the amount to which the CMB 
analysis may be compromised by the residual SZ component, for which we 
need analysis as model independent as possible.  

Our principal method, described in section \ref{sec:est}, constructs a linear 
combination of the \WMAP\ Q, V, and W band cross-power spectra which maximizes 
the contribution of SZ, while at the 
same time minimizing the radio point source and CMB contribution.  We use 
this linear combination to fit for the SZ power spectrum amplitude, 
using a spectrum shape from halo model calculations \cite{2002MNRAS.336.1256K}.
 
In section \ref{sec:markov} we supplement this method by 
fitting for cosmological parameters from the \WMAP\ temperature power spectrum 
using the Markov chain Monte Carlo method, allowing for an additional 
SZ component in the power spectrum with a free amplitude to be determined by 
the 
data.  
The two methods give consistent results, with the latter giving 
somewhat weaker constraints. Conclusions are presented in section 
\ref{sec:conclusion}. 

\section{Cross-spectra combination for SZ} \label{sec:est}
We begin our discussion with a simple
 example.  In the absence of noise and point sources, the exact SZ power 
spectrum could be computed from the difference of any two cross-spectra from 
different bands.  Suppose we took the differencing assemblies from the Q, V and
 W bands, and compute the cross-power spectra. The SZ power is given by, for 
example, the difference $($Q1V1$)_l-($W1W2$)_l = 0.25\ C_l^\text{SZ,RJ}$ in 
this idealize case, where the coefficient is computed from table 
\ref{tab:wmapbands}.  Other combinations would also yield the SZ spectrum, with
 a different coefficient. The 
important things to note is that the coefficient is not very small and 
the CMB cancels exactly in this combination: any CMB cosmic variation in the Q 
and V bands is the same as in the W band, and is subtracted out.  In the 
presence of instrument noise, this difference does not give the exact SZ power 
spectrum, but a noisy estimate for it.  Given a power spectrum shape, we may 
estimate its amplitude by summing together different $l$ bins, each weighted to
give the proper normalization and to emphasize bins with high signal to noise. 
Our SZ estimator works in this fashion, except that in the presence of point 
sources, QV$-$WW is a biased estimate.  However, by including other 
cross-spectra, we can form a linear combination of the cross-spectra bins to 
yield an unbiased estimate for the SZ amplitude.  In the following we find such
a combination, and apply it to \WMAP\ data.  

\subsection{Estimator and data} 
In this section, we introduce our estimator and list the data we use.  Our 
estimator, derived in appendix \ref{sec:estderiv}, is a generalized version of 
the point source estimator of \citeN{2003ApJS..148..135H}.

We want to generate an estimate for SZ from the \WMAP\ cross-spectra. We 
postulate that the data is the sum of the CMB and two contaminants, radio point
sources and SZ.  We marginalize point sources and estimate SZ in our main case,
which we illustrate in detail below.  We consider other cases also, but for 
brevity omit their details, which are similar.
We can write the cross-spectra as
\be
\langle  C^\bi{i}_l \rangle = C^{\bi{i},\rm CMB}_l +  C^{\bi{i},\rm src}_l + 
C^{\bi{i},\rm SZ}_l,
\ee
showing explicitly the contribution from each part of the signal.  Here the 
multipole bin is denoted by $l$ and the pair of differencing assemblies in the 
cross-correlation is denoted by $\bi{i} = {i_1 i_2} = $ W1W2, Q1V1, {\it etc.} 
No auto-power spectra are included, so we need not worry about noise 
subtraction. 

We will marginalize over the CMB spectrum, which we denote by $C_l^\text{CMB}$.
 The window functions for each differencing assembly pair are $\beps = \{ 
w^\bi{i}_{ll'} \}$.  We boldface $\beps$ because later we will think of it as a
matrix.  Therefore the contribution to the cross-spectrum from the CMB is
\be
C^{\bi{i},\rm CMB}_l =  \sum_l w^\bi{i}_{ll'} C^\text{CMB}_{l'}.
\ee
The spectra we use are in temperature units, and have already been 
beam-deconvolved, so the window functions $w^\bi{i}_{ll'} = \delta_{ll'}$ are 
trivial.  However, it is necessary to keep the window functions explicit in the
manipulations that follow.

We denote the amplitude of the point source power spectrum by $A$. This 
amplitude relates to the cross-spectra via the frequency and shape dependence 
$\bS = \{  S^\bi{i}_{l} \}$.  Later we will think of $\bS$ as a vector.  We 
take radio sources to have a white noise spectrum with power law frequency 
dependence, given by:
\bea\nonumber
C_l^{\bi{i},\rm src} &=& S^\bi{i}_l A \\ 
 S^\bi{i}_{l} &=& \left( \frac{\nu_{i_1}}{ \nu_0} \right)^{\beta} \left( 
\frac{\nu_{i_2} }{ \nu_0} \right)^{\beta},
\eea 
where cross-spectrum $\bi{i}$ has channels at $\nu_{i_1}$ and $\nu_{i_2}$.  The
units of $A$ are temperature squared.  Well-resolved point sources have already
been masked from the maps before the evaluation of the cross-spectra, so $A$ 
represents unresolved sources only. \citeN{2003ApJS..148...97B} found 
$\beta=-2.0$ and $\nu_0=45$ GHz for the resolved sources.  Following 
\citeN{2003ApJS..148..135H} we take the same for the unresolved sources.  

We describe amplitude of SZ with $B$, the ratio of the SZ power spectrum and 
the predicted spectrum, assuming they have the same shape.  Thus the SZ 
amplitude is dimensionless, and has a theoretically predicted value $B=1$ for 
$\sigma_8=0.9$ using the halo models of \citeN{2002MNRAS.336.1256K}.  
In our notation, this amplitude also relates to the power spectra by the 
frequency dependence and shape for SZ, which we label $\Z = \{  Z^\bi{i} _{l} 
\}$.  
The frequency dependence of a temperature perturbation due to SZ is
\be
\frac{\Delta T^\text{SZ}}{ T^\text{CMB}} \propto -2 f(x),
\ee
where the frequency dependence relative to the RJ regime is given by\be
f(x) = 2 - \frac{x/2}{\tanh(x/2)}, \label{eqn:szfreqdep}
\ee
where $x=h\nu/k_B T_\text{CMB}$.  Note $f \rightarrow 1$ in the RJ limit $x 
\rightarrow 0$.
%At a given frequency, SZ is proportional to the comptonization parameter 
%$y(\n)$, given by the integral of the electron pressure along the line of 
%sight: 
%\begin{equation}
%  y(\n) = \int \sigma_T n_e(\n) {k_B T_e(\n) \over m_e c^2} a d\chi.
%\end{equation}
%The integral is over the comoving radial coordinate $\chi$.  Here $\sigma_T$ 
%is the Thomson scattering cross-section, $n_e$ is the electron density, and 
%$T_e$ is the electron temperature.  
Thus the SZ contribution to the cross-spectrum is:
\bea \label{eqn:szpow} \nonumber
C_l^{\bi{i},\rm SZ} &=& Z^\bi{i}_l B \\
Z^\bi{i}_l &=& C_l^\text{SZ,RJ} f(\nu_{i_1})f(\nu_{i_2}),
\eea
where the spectrum shape $C_l^\text{SZ,RJ}$ is from the halo model prediction 
in the RJ regime using 
$\sigma_8 = 0.9$, as shown in Figure \ref{fig:limit}.  The shape of SZ is 
roughly $C_l^\text{SZ,RJ} \propto l^{-1}$, although it becomes slightly steeper
for $l$ greater than a few hundred. 

We organize the binned cross-spectra $C^\bi{i}_l$ into a data vector $\D = \{ 
C^\bi{i}_l \}$.   We use a Gaussian model for the likelihood $\like$ of the 
power spectrum:
\be
-2 \log \like \propto \left[ \D -  \langle \D \rangle\right]\t \Sgi \left[ \D -
 \langle \D \rangle \right].
\ee
where the covariance $\Sg = \langle (\D-\langle \D \rangle)(\D-\langle \D 
\rangle)\t \rangle$ can be written as $\Sg = \{ \Sigma^{\bi{ii'}}_{ll'} \}$.  
We derive $\bar B$, an unbiased estimator for $B$, and its covariance 
$\Sigma^B$ in appendix \ref{sec:estderiv}.  The main result is:
\bea \label{eqn:est}\nonumber 
\bar B &\equiv& (\Z\t\F\Z)^{-1} \Z\t \F \D \\ 
\Sigma^B &\equiv& (\Z\t \F \Z)^{-1},
\eea
where we have defined the auxiliary matrices
\bea \label{eqn:estaux}\nonumber
\F &\equiv& \E - \E \bS\left(\bS\t \E \bS\right)^{-1}\bS\t \E \\ 
\E &\equiv& \Sgi -  \Sgi \beps\left(\beps\t \Sgi \beps\right)^{-1} \beps\t \Sgi
\eea
In this notation, we consider $\D$, $\bS$, and $\Z$ as column vectors with 
single index $\bi{i}l$, and $\beps$ as matrix with indices $\bi{i}l$ and $l'$. 
$\Sg$, $\E$, and $\F$ are matrices with indices $\bi{i}l$ and $\bi{i}'l'$. This
estimator marginalizes CMB and point sources, which is the most conservative 
treatment, since it assumes we know nothing about these components.  
A more aggressive treatment would be to estimate all 3 components 
simultaneously, but we defer this approach to a future analysis. 
Note that $\bar B$ is a linear combination of the cross-spectra $\D$, with 
weights $(\Z\t\F\Z)^{-1} \Z\t \F$.

The data we use consist of the 28 cross-power spectra from the eight 
differencing assemblies in the \WMAP\ Q, V, and W bands.  These spectra are 
provided at the {\em Legacy Archive for Microwave Background Data Analysis} 
[\onlinecite{lamda_cross_powspec}].  A galactic foreground model 
\cite{2003ApJS..148...97B} has already  been  subtracted out.  \WMAP's 
temperature power spectrum is a linear combination of these 28 cross-spectra 
\cite{2003ApJS..148..135H}.  

We bin the spectra in $l$, accounting for the number of modes at each $l$.  The
{\em Legacy Archive} does not provide the cross-spectrum covariance, which we 
need for $\Sg$ in equation (\ref{eqn:estaux}), so we estimate the covariance 
from the data.  We assume the covariance is diagonal in $l$, which is a good 
approximation \cite{2003ApJS..148..135H}.  Then in a single bin, we use the 
dispersion of the cross-spectra about \WMAP's combined temperature power 
spectrum to estimate $\Sg$ for that bin.  (For this purpose we must first 
un-correct the combined spectrum for the point source contribution.)  This 
procedure to obtain the covariance works best if the power spectrum variance 
does not change much within a bin and if the bin contains enough $l$'s to get a
low-noise estimate of the variance.  Our bin width of $\Delta l=40$ is fairly 
narrow, and gives a covariance which is numerically stable in our subsequent 
calculations.  This technique does not account for the cosmic variance of the 
CMB, but as we note in the derivation in the appendix, our estimator is 
insensitive to this source of variance.  This makes sense because the CMB is 
completely projected out, as in our QV$-$WW example at the beginning of section
\ref{sec:est}.

As a test of our estimator, we repeat the estimate of 
\citeN{2003ApJS..148..135H} 
of the power spectrum amplitude of unresolved point sources.  In this case we 
neither marginalize nor estimate SZ, but assume $B=0$ in equation 
(\ref{eqn:szpow}).  We find a point source amplitude of $A= 0.016 \pm 0.001\ 
\mu\text{K}^2$, which is roughly consistent with their quoted value of 
$A=0.015\ \mu\text{K}^2$.  This gives us confidence in our estimator, despite 
our covariance matrix constructed from the data.

\subsection{Results} \label{sec:result}
\begin{figure}
\begin{center}
\includegraphics[width=\figwidth]{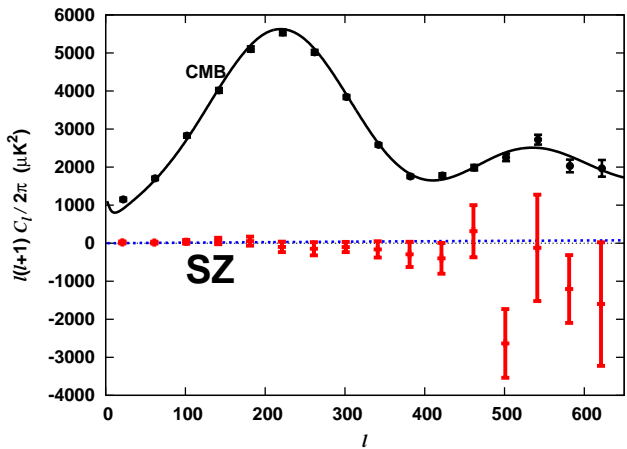}
\end{center}
\caption{Reconstruction of the SZ power spectrum in the RJ regime,
 in bins of $\Delta l = 40$ 
(red error bars).  There is no visible 
SZ detection, which would have been positive for this combination. 
For comparison, we have included the \WMAP\ combined temperature power 
spectrum, binned the same way, the \WMAP\ best-fit theoretical CMB power 
spectrum, and 
a theoretical prediction of SZ in the RJ regime for $\sigma_8 = 0.9$ (blue 
dotted), calculated from a halo model.} \label{fig:szestpow}
\end{figure}

We show our estimates for SZ on a bin-by-bin basis, before combining different 
multipole bins to improve statistics.  We marginalize over the CMB and the 
point source amplitude.  Our best estimate for the binned SZ power spectrum is 
shown in Figure \ref{fig:szestpow}.  Immediately we can see that the data do 
not tolerate a large SZ contribution.

\begin{figure}
\begin{center}
\includegraphics[width=\figwidth]{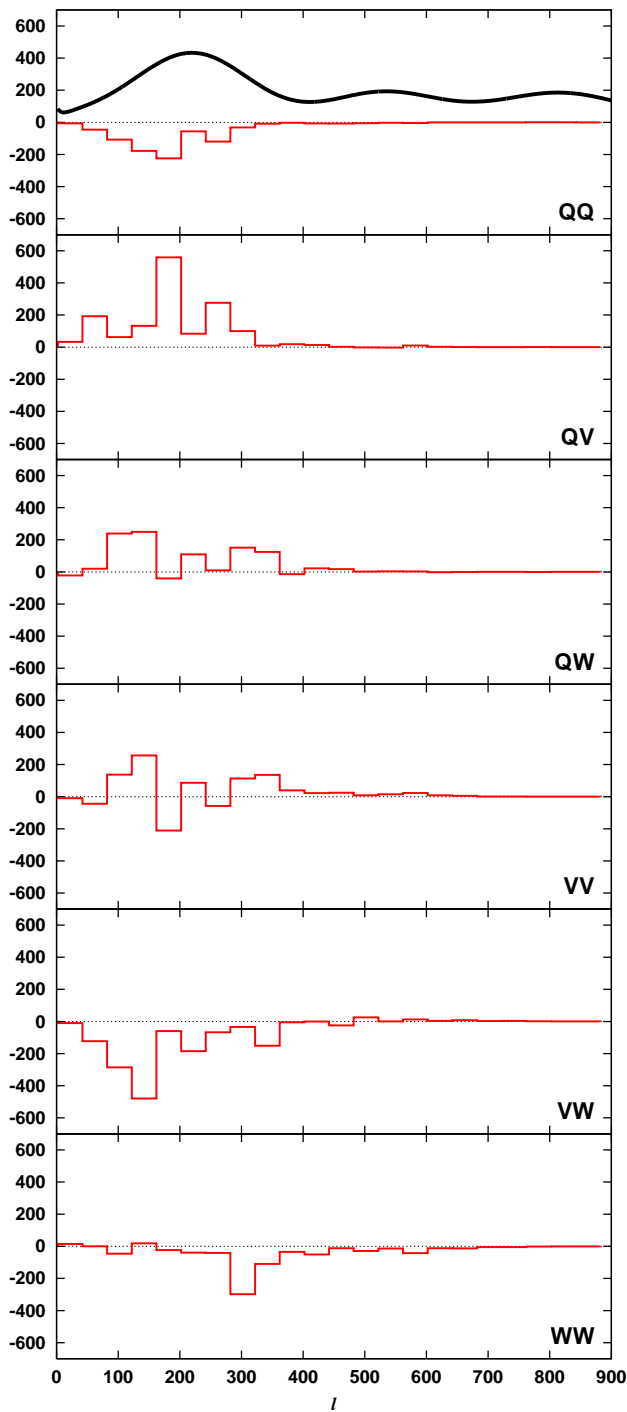}
\end{center}
\caption{The contribution to the SZ amplitude estimator $\bar B$ from the 
cross-spectra.  The horizontal axis is the multipole $l$.  The vertical axis 
show the weight of each spectrum in $\mu$K$^{-2}$.  We have grouped the 
cross-spectra by band as noted.  Within each group, we have summed the weights 
of the individual spectra.  In the top frame, we have included the CMB power 
spectrum (arbitrary units) to give scale for $l$.} \label{fig:weight}
\end{figure}

Next we turn to our main case, where we combine all the bins together and 
estimate the SZ amplitude. Figure \ref{fig:weight} shows the weight 
$(\Z\t\F\Z)^{-1} \Z\t \F$ for each cross-spectrum in the amplitude estimator.  
The weights have been divided into groups based on the bands.  For plotting, we
sum the weights in each group.  The bulk of the weight comes from $l=100$ to 
$l=400$.  At lower $l$ there are fewer modes and the SZ contribution 
is low.  At higher $l$ the statistics are limited by detector noise. 

The weights are difficult to interpret heuristically.  From the frequency 
dependence, the strength of SZ decreases in order of QQ, QV, VV, QW, VW, and 
WW.
 We would expect our combination of cross-spectra would have the SZ-stronger 
bands minus the SZ-weaker bands.  So it is easy to understand the positive 
weights of QV and QW and the negative weights of VW and WW.  Point sources are 
strongest in Q-band, so a negative weight for QQ makes sense in terms of a 
point source correction.  QW and VW are positive except for bins which are 
strong in QV.  These negative bins may also represent a point source correction

\begin{figure}
\begin{center}
\includegraphics[width=\figwidth]{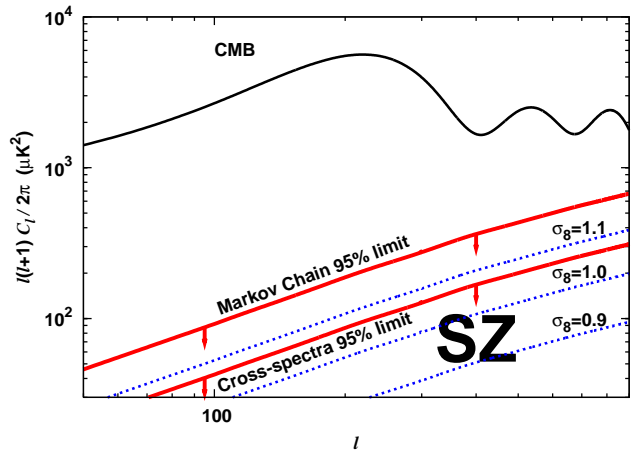}
\end{center}
\caption{Limit on SZ power spectrum.  We present a 95\% upper limit on the 
amplitude of the SZ power spectrum from a combination of WMAP band 
cross-spectra.  We show the temperature fluctuation in the Rayleigh-Jeans 
regime.  For comparison we display the \WMAP\ best fit power spectrum and SZ 
power spectra from a halo model calculation using $\sigma_8=0.9, \, 1, \, 
1.1$.}\label{fig:limit}
\end{figure}

For the amplitude of SZ relative to the expected RJ amplitude from the halo 
model with \WMAP\ parameters, our estimator gives us $\bar B = -0.042 \pm 
1.685$.  The error is large, and is consistent with both no SZ and the expected
$B=1$.  Including only the physical region $B>0$, we integrate the likelihood 
until we include 95\% of the probability.  From this we quote an upper limit of
\be
{B} < 3.3 \quad (95\%). \label{eqn:limit}
\ee
We plot this limit, along with some models of the SZ power spectrum in Figure 
\ref{fig:limit}.  At the first acoustic peak, SZ is less than 3\% of the CMB in
the RJ regime and less than 2\% of CMB in W band, on which 
cosmological parameter estimation is heavily based.  
% In the V and W bands, on which the cosmological parameter estimation is 
%heavily based, the contamination is less than 4\% and 3\% respectively.  
We find no evidence for a large SZ contribution.  Assuming $C_l^\text{SZ} 
\propto (\sigma_8)^7$, this gives a limit of $\sigma_8 < 1.07$ at 95\% 
confidence.  To this one should add an additional 
modeling uncertainty at the level of 
10\% based on the comparison of predictions
from different simulations \cite{2002MNRAS.336.1256K}.
There is a caveat in the upper limit derived here in that we are 
working with power spectra based on masked maps in \WMAP, with 
more than 200 radio point sources removed. If the SZ signal is correlated with 
these point sources, which could happen if these radio sources sit in 
massive clusters \cite{2002ApJ...580...36H}, then more 
SZ may have been removed  than expected based on the sky fraction of the mask. 
This would only affect the $\sigma_8$ limits and not the SZ contamination 
on the primary CMB in the \WMAP\ power spectra. 
Since we are mostly concerned with the latter 
we do not explore this issue  further. 
Note that the upper limit on $\sigma_8$ 
is already 
comparable to the predicted value based on detections from CBI and BIMA, 
which gives $\sigma_8 \sim 0.95-1.05$
\cite{2004astro.ph..2359R,2002MNRAS.336.1256K}. It is remarkable that \WMAP\ 
first year data 
have sufficient sensitivity to place constraints on the SZ amplitude 
comparable to other small scale surveys. 

\begin{figure}
\begin{center}
\includegraphics[width=\figwidth]{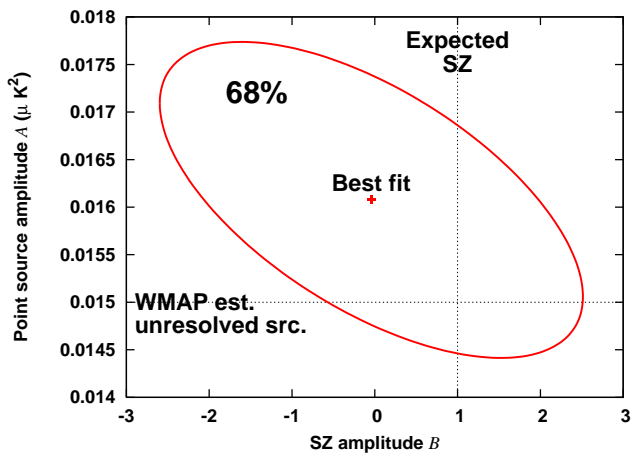}
\end{center}
\caption{
A joint estimation of the SZ power spectrum amplitude and the point source 
spectrum amplitude.  The ellipse encloses 68\% of the likelihood.  The best fit
point is shown by a ``$+$''.  The vertical line at B=1 shows the theoretically 
predicted SZ amplitude.  The horizontal line shows \WMAP's value for point 
sources, which does not take into account SZ.  Because the power spectrum is a 
positive quantity, note that portions of the plane where either axis is 
negative are unphysical.
} \label{fig:sz-and-pnt}
\end{figure}

In the next application 
we jointly estimate the point source and SZ amplitude, rather than directly 
marginalizing over the point sources. We find the two 
parameters to be somewhat correlated (Figure \ref{fig:sz-and-pnt}), but not to 
the point of allowing very large SZ amplitude.   An extremely strong SZ power 
is not allowed. 
%Projecting over the point sources in this plane gives constraints on SZ 
%amplitude comparable to those in equation \ref{eqn:limit}.

\section{Markov chain analysis} \label{sec:markov}

In this section we estimate how much SZ signal is in the \WMAP\ $C_l$ power 
spectrum, and investigate
 the effect of SZ on the determination of the cosmological parameters.  For 
this purpose we use the Markov Chain Monte Carlo (MCMC) approach, using 
software described in more detail elsewhere 
\cite{2003MNRAS.342L..79S,2004astro.ph..3073S}.

We ran two MCMCs, one without SZ and one allowing for an unconstrained SZ
contribution.  We built a third chain from the second by importance sampling, 
allowing for an SZ component but constraining it to limits derived
based on frequency information in the previous section.
We used the \WMAP\ likelihood routine 
\cite{2003ApJS..148..135H,2003ApJS..148..195V}.   Each of the chains contains 
100,000 total chain elements.  The success rate is 45--55 percent, the 
correlation length is 13--20 elements, and the effective length is 
5,000--10,000 elements.  Each chain comprises 23 independent sub-chains and, in
terms of Gelman and Rubin $\hat R$-statistics \cite{gelman92}, we find the 
chains are sufficiently converged and mixed ($\hat R<1.01$, compared to the 
recommended value of $\hat R<1.2$).

In the second chain, we added to the power spectrum an SZ-shaped contribution, 
parameterized in terms of amplitude $B$ (equation \ref{eqn:szpow}). The \WMAP\ 
power spectrum combines the Q, V and W bands in different ratios at each $l$, 
so the shape of the SZ contribution to the \WMAP\ power spectrum is not exactly
given by 
the single frequency SZ template, because the effective frequency for every $l$
varies.  This dependence is small and we ignore it here. We find that the SZ 
contribution to the \WMAP\ combined temperature power spectrum, dominated by V 
and W channels, may be approximated as 75\% the contribution in RJ.  One could 
also add additional CMB experiments (e.g. CBI, VSA, {\it etc.}) into the 
analysis, but this 
would incur complications to account for the different frequencies of these 
experiments. 
In the third chain we add our multi-frequency analysis limit as an additional
constraint. 

We consider only the simplest model required by the data plus the SZ component,
since we want to analyze the effect of the latter on the cosmological 
parameters.  \WMAP\ temperature data require neither tensor modes nor curvature
nor 
running of the primordial power spectrum of the scalar perturbations, so we do 
not consider them.  

We work in a seven parameter space:
\be
{\bi p}  = \{ \omega_b,\ \omega_\text{cdm},\ \Omega_m,\ \tau,\ A_s,\ n_s,\ 
B \}.
\label{eq:parameterization}
\ee
Here $\omega_b = \Omega_b h^2$ is the baryonic content of the universe, 
$\omega_\text{cdm} = \Omega_\text{cdm} h^2$ is the physical density of the  
cold dark matter content, 
$\Omega_m=\Omega_\text{cdm}+\Omega_b=1-\Omega_{\Lambda}$ is 
the matter density today, $\tau$ is the optical depth to reionization, $A_s$ is
the amplitude of the primordial scalar perturbations, $n_s$ is the primordial 
slope, and as before $B$ is the amplitude of the SZ power spectrum.

To reduce the degeneracies while running the MCMCs,
we use $\omega_b$, $\omega_\text{cdm}$, angular size of the sound horizon 
$\Theta_s$, $\lg {A_s}$, $n_s$, $\lg {A_s}-\tau - 0.5\lg 
(\omega_b+\omega_\text{cdm})$, and $B$, instead of the parameters in equation 
\ref{eq:parameterization}.  We adopt broad flat priors on these parameters, and
additionally require $\tau<0.3$.

We find that the amplitude of an SZ-shaped component to the \WMAP\ power 
spectrum is limited to $B<7.1$ at 95 percent confidence  (see Figure 
\ref{fig:limit}).  This limit means that the contribution to the \WMAP\ 
temperature power spectrum at the first peak is below 5 percent.  This is a 
weaker limit than from the combination of cross-spectra, where a factor of 2 
better limit was found (equation 
\ref{eqn:limit}).

\begin{table*}
\noindent
\begin{center}
\begin{ruledtabular}
\begin{tabular}{lrrr}
%\hline \hline \\
& no SZ & with SZ, no SZ prior & with SZ prior \\ \hline
% & & &  \\
$B$ & 
0  & $<7.1$ (95\%)& 
$<2.9$  (95\%)
\\ %& &  & \\
$\omega_b \times 10^2$ & 
$2.35^{+0.14}_{-0.13}\;{}^{+0.28}_{-0.26}$&
$2.51^{+0.21}_{-0.18}\;{}^{+0.47}_{-0.33}$&
$2.43^{+0.15}_{-0.15}\;{}^{+0.30}_{-0.29}$
\\ %& & &  \\
$\Omega_m$ &
$0.245^{+0.07}_{-0.06}\;{}^{+0.15}_{-0.10}$ &
$0.234^{+0.07}_{-0.06}\;{}^{+0.15}_{-0.10}$&
$0.243^{+0.07}_{-0.06}\;{}^{+0.15}_{-0.10}$
\\ %& & &  \\
$\omega_\text{cdm}$ & 
$0.111^{+0.016}_{-0.015}\;{}^{+0.033}_{-0.029}$ &
$0.111^{+0.016}_{-0.016}\;{}^{+0.033}_{-0.030}$&
$0.111^{+0.016}_{-0.016}\;{}^{+0.033}_{-0.031}$
\\ %& & &  \\
$\tau$ & 
$0.19^{+0.07}_{-0.08}\;{}^{+0.10}_{-0.14}$ &
$0.20^{+0.07}_{-0.08}\;{}^{+0.10}_{-0.14}$ &
$0.19^{+0.07}_{-0.08}\;{}^{+0.10}_{-0.14}$
\\ %& & &  \\
$\sigma_8$ & 
$0.88^{+0.12}_{-0.11}\;{}^{+0.25}_{-0.20}$&
$0.86^{+0.12}_{-0.11}\;{}^{+0.24}_{-0.22}$ &
$0.87^{+0.12}_{-0.11}\;{}^{+0.23}_{-0.22}$
\\ %& & &  \\
$h$ & 
$0.74^{+0.06}_{-0.05}\;{}^{+0.12}_{-0.09}$ &
$0.76^{+0.07}_{-0.06}\;{}^{+0.14}_{-0.11}$&
$0.75^{+0.06}_{-0.05}\;{}^{+0.13}_{-0.10}$
\\ %& & &  \\
$n_s$ &
$0.99^{+0.04}_{-0.04}\;{}^{+0.07}_{-0.07}$ &
$0.99^{+0.04}_{-0.04}\;{}^{+0.07}_{-0.07}$ &
$0.99^{+0.04}_{-0.04}\;{}^{+0.07}_{-0.07}$
\\ %& & & \\
%\hline \hline
\end{tabular}
\end{ruledtabular}
\end{center}
\caption{ 
The first two columns contain the median value and  1- and $2\sigma$ 
constraints on cosmological parameters for two MCMCs without and with a 
Sunyaev-Zeldovich component in the $C_l$ power spectrum from \WMAP\ data alone.
For both chains there was an imposed prior of $\tau<0.3$. The third column 
shows the constraints when the limit on SZ from our cross-spectrum estimator is
applied as a prior.\label{table:SZparameters}
}
\end{table*}

\begin{figure}
\begin{center}
\includegraphics[width=\figwidth]{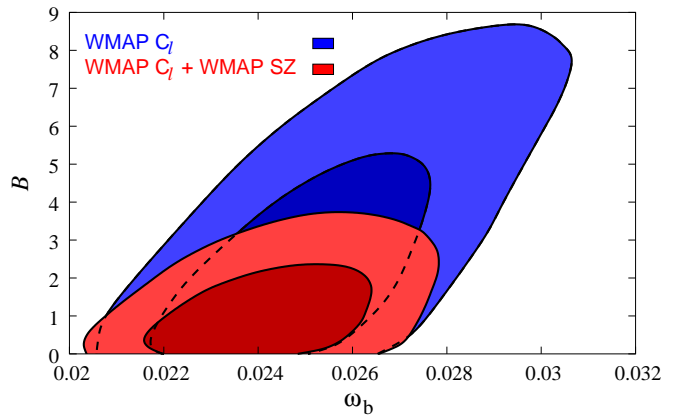}
\end{center}
\caption{
We show two-dimensional contours of 68\% and 95\% confidence levels in 
$(\omega_b,B)$ plane.  The larger (blue) contours show the degeneracy using the
\WMAP\ combined temperature power spectrum without any prior on SZ.  
The smaller (red) contours include our multi-frequency cross-spectrum analysis 
as a prior.  $B=1$ is expected from halo models for SZ when $\sigma_8=0.9$.
} 
\label{fig:omegabSZ}
\end{figure}

Table~\ref{table:SZparameters} shows the comparison of the two MCMCs, showing 
the effect of including SZ in the analysis. 
The inclusion of the additional parameter $B$ into
the likelihood analysis affects only the determination of 
the baryon physical density $\omega_b$.  Without SZ we find $\omega_b=0.0235$,
whereas with SZ we find $\omega_b=0.0251$, which is shifted
by about $1.5\sigma$ away from the earlier value. The confidence
contours in the $(\omega_b,B)$ plane are shown in
Figure~\ref{fig:omegabSZ}, showing that there is a degeneracy
between these two parameters.
However, we can and should also 
use the constraint from our multi-frequency cross-spectrum analysis as a prior.
 We can include the Gaussian likelihood for $B$ from section \ref{sec:est}, and
perform importance sampling of the chain with SZ.  We then find $\omega_b = 
0.0243$, different from the case without SZ by $0.6\sigma$.  The likelihood 
contours including the prior are also shown in Figure~\ref{fig:omegabSZ}. The 
other parameters are much less affected by the SZ. 
The next largest deviation from the first chain to the second is a shift in the
median $h$ of about $0.3\sigma$ towards higher value, which is not 
statistically significant and the effect is even smaller if the constraints 
from previous 
section are added to the chains.

\section{Conclusions} \label{sec:conclusion}

We have analyzed the \WMAP\ power spectra in two ways and found them to be free
of serious contamination from SZ.
The more powerful and less model dependent 
of the two methods is to use multi-frequency information: 
by combining the \WMAP\ cross-spectra, we limit at 95\% confidence the 
amplitude of SZ to be below 2\% of the CMB at the position of first peak in W 
band.
By searching for an SZ-shaped component 
in the \WMAP\ combined spectrum with a Markov chain Monte Carlo we also 
do not find any evidence of a signal, but we can only set a weaker limit.  
Combining the analyses, we show that the cosmological parameters are not 
affected by the SZ within the range allowed by the multi-frequency analysis.

There are several possible improvements in the analysis that could 
further tighten
the limits and may be applied to the second year data once they 
become available. 
In addition to having lower noise, the two year data are also expected to have 
better control of systematics such as beam uncertainties  and noise 
correlations.

First, we would like to improve our method for obtaining the cross-spectrum 
covariance matrix.  The estimate for the covariance is noisy, which makes our 
weights for combining the power spectra noisier than they should be.  If we 
double the size of the bins we average over to obtain the covariance, our limit
changes slightly ($B<4.3$ opposed to $B<3.3$ at 95\%) and the point source 
estimate is virtually unchanged.  However, it is impossible to tell if this 
effect is due to a covariance matrix with less noise (which is better) or due 
to weights whose wide bins combine high- and low-signal-to-noise multipoles 
less optimally (which is worse).  
This can be improved if the \WMAP\ second year data release includes the 
cross-spectrum covariance matrix.

The second possible 
improvement of our analysis  is to improve
the frequency dependence of point sources or the \WMAP\ beam deconvolution, 
both of which can bias our estimate.  Our estimator is only unbiased if the 
models one applies to it are faithful to the data.  The models we have used are
the best available, and this information will improve in the second year data.
It seems unlikely that these improvements
will strongly modify the conclusions in this 
paper. 

The third improvement is to perform the analysis on the maps which do not 
have point sources excised. As discussed above,  
masking of resolved point sources from the \WMAP\ data may reduce the SZ signal
if the two are correlated. This would weaken our limit on $\sigma_8$ 
from the absence of SZ signal,
but does not change our conclusions about the effect of SZ on 
the CMB and cosmological parameters.
Still, given that the current limits are reaching the levels of interest 
it is worth exploring this possibility further. 

The fourth improvement is to verify the assumed 
shape for the SZ power spectrum. This is currently based on halo models, 
since hydrodynamic simulations cannot simulate sufficiently large volumes
to make predictions reliable. 
 If the power spectrum has a radically different shape, our limit is hard to 
interpret, but the level of contamination on primary CMB is likely to remain 
unchanged, as is clear from  Figure \ref{fig:szestpow}.  As an example we
estimated the SZ in bins of $\Delta l=120$.  In the bin containing the first 
acoustic peak, we limit the SZ power to 4.9 times the halo model power 
predictions at 95\% confidence or 4\% contamination in W. 
It is clear that as long as the SZ spectrum is smooth
the limits on the contamination are not going to change significantly over
the range with best sensitivity at $100<l<300$. 

Finally, our procedure to marginalize over CMB and 
point sources is conservative 
and the errors could be further reduced if a joint estimation is used 
instead. 
We know something more about the CMB power spectrum than just its
frequency dependence, and in this case we can use this information.
We also note that a relatively trivial modification to \WMAP's method for 
estimating the power spectrum could eliminate any contaminant with an SZ-like 
shape and frequency dependence, much in the same way that they eliminate point 
sources.
However, the fact that we find no residual SZ 
effects implies that this procedure is not 
really required and justifies the treatment of the \WMAP\ team in ignoring SZ. 

Even with future data detecting the SZ power spectrum from \WMAP\ alone 
may be difficult.  If the bulk of the covariance is from noise, it would take 
\WMAP\ roughly 3 years to get a 1$\sigma$ detection of SZ using the 
cross-spectrum combination method of this paper, if $\sigma_8=0.9$.  
Elimination of non-noise sources of covariance, such as uncertainties remaining
in the beams, may help.
If $\sigma_8=1$ then a 2$\sigma$ detection should be possible with 
4 year data and one should start seeing a hint of the signal already with 
2 year data. Theoretical predictions remain somewhat uncertain, so it is 
possible that these predictions have a systematic uncertainty of around 
10\%, but 
it is clear that it will be worth looking for SZ in future WMAP releases. 

\section*{Acknowledgments}
KMH wishes to thank Chris Hirata and Nikhil Padmanabhan for enlightening 
discussions.
We thank Eiichiro Komatsu for his assistance with the model SZ.
We acknowledge the use of the Legacy Archive for Microwave Background Data 
Analysis (LAMBDA). Support for LAMBDA is provided by the NASA Office of Space 
Science.
Our MCMC simulations were run on a Beowulf cluster at Princeton University,
supported in part by NSF grant AST-0216105.
US is supported by Packard Foundation, Sloan Foundation,
NASA NAG5-1993 and NSF CAREER-0132953.

%\section*{References}
\bibliography{cosmo,cosmo_preprints,sz_limit} %\bibliographystyle{mnras}

\begin{thebibliography}{24}
\expandafter\ifx\csname natexlab\endcsname\relax\def\natexlab#1{#1}\fi
\expandafter\ifx\csname bibnamefont\endcsname\relax
  \def\bibnamefont#1{#1}\fi
\expandafter\ifx\csname bibfnamefont\endcsname\relax
  \def\bibfnamefont#1{#1}\fi
\expandafter\ifx\csname citenamefont\endcsname\relax
  \def\citenamefont#1{#1}\fi
\expandafter\ifx\csname url\endcsname\relax
  \def\url#1{\texttt{#1}}\fi
\expandafter\ifx\csname urlprefix\endcsname\relax\def\urlprefix{URL }\fi
\providecommand{\bibinfo}[2]{#2}
\providecommand{\eprint}[2][]{\url{#2}}

\bibitem[{\citenamefont{{Bennett}
  et~al.}(2003{\natexlab{a}})\citenamefont{{Bennett}, {Halpern}, {Hinshaw},
  {Jarosik}, {Kogut}, {Limon}, {Meyer}, {Page}, {Spergel}, {Tucker}
  et~al.}}]{2003ApJS..148....1B}
\bibinfo{author}{\bibfnamefont{C.~L.} \bibnamefont{{Bennett}}},
  \bibinfo{author}{\bibfnamefont{M.}~\bibnamefont{{Halpern}}},
  \bibinfo{author}{\bibfnamefont{G.}~\bibnamefont{{Hinshaw}}},
  \bibinfo{author}{\bibfnamefont{N.}~\bibnamefont{{Jarosik}}},
  \bibinfo{author}{\bibfnamefont{A.}~\bibnamefont{{Kogut}}},
  \bibinfo{author}{\bibfnamefont{M.}~\bibnamefont{{Limon}}},
  \bibinfo{author}{\bibfnamefont{S.~S.} \bibnamefont{{Meyer}}},
  \bibinfo{author}{\bibfnamefont{L.}~\bibnamefont{{Page}}},
  \bibinfo{author}{\bibfnamefont{D.~N.} \bibnamefont{{Spergel}}},
  \bibinfo{author}{\bibfnamefont{G.~S.} \bibnamefont{{Tucker}}},
  \bibnamefont{et~al.}, \bibinfo{journal}{\apjs}
  \textbf{\bibinfo{volume}{148}}, \bibinfo{pages}{1}
  (\bibinfo{year}{2003}{\natexlab{a}}).

\bibitem[{\citenamefont{{Bennett}
  et~al.}(2003{\natexlab{b}})\citenamefont{{Bennett}, {Hill}, {Hinshaw},
  {Nolta}, {Odegard}, {Page}, {Spergel}, {Weiland}, {Wright}, {Halpern}
  et~al.}}]{2003ApJS..148...97B}
\bibinfo{author}{\bibfnamefont{C.~L.} \bibnamefont{{Bennett}}},
  \bibinfo{author}{\bibfnamefont{R.~S.} \bibnamefont{{Hill}}},
  \bibinfo{author}{\bibfnamefont{G.}~\bibnamefont{{Hinshaw}}},
  \bibinfo{author}{\bibfnamefont{M.~R.} \bibnamefont{{Nolta}}},
  \bibinfo{author}{\bibfnamefont{N.}~\bibnamefont{{Odegard}}},
  \bibinfo{author}{\bibfnamefont{L.}~\bibnamefont{{Page}}},
  \bibinfo{author}{\bibfnamefont{D.~N.} \bibnamefont{{Spergel}}},
  \bibinfo{author}{\bibfnamefont{J.~L.} \bibnamefont{{Weiland}}},
  \bibinfo{author}{\bibfnamefont{E.~L.} \bibnamefont{{Wright}}},
  \bibinfo{author}{\bibfnamefont{M.}~\bibnamefont{{Halpern}}},
  \bibnamefont{et~al.}, \bibinfo{journal}{\apjs}
  \textbf{\bibinfo{volume}{148}}, \bibinfo{pages}{97}
  (\bibinfo{year}{2003}{\natexlab{b}}).

\bibitem[{\citenamefont{{Ebeling} et~al.}(1996)\citenamefont{{Ebeling},
  {Voges}, {Bohringer}, {Edge}, {Huchra}, and {Briel}}}]{1996MNRAS.281..799E}
\bibinfo{author}{\bibfnamefont{H.}~\bibnamefont{{Ebeling}}},
  \bibinfo{author}{\bibfnamefont{W.}~\bibnamefont{{Voges}}},
  \bibinfo{author}{\bibfnamefont{H.}~\bibnamefont{{Bohringer}}},
  \bibinfo{author}{\bibfnamefont{A.~C.} \bibnamefont{{Edge}}},
  \bibinfo{author}{\bibfnamefont{J.~P.} \bibnamefont{{Huchra}}},
  \bibnamefont{and} \bibinfo{author}{\bibfnamefont{U.~G.}
  \bibnamefont{{Briel}}}, \bibinfo{journal}{\mnras}
  \textbf{\bibinfo{volume}{281}}, \bibinfo{pages}{799} (\bibinfo{year}{1996}).

\bibitem[{\citenamefont{{Afshordi} et~al.}(2003)\citenamefont{{Afshordi},
  {Loh}, and {Strauss}}}]{2003astro.ph..8260A}
\bibinfo{author}{\bibfnamefont{N.}~\bibnamefont{{Afshordi}}},
  \bibinfo{author}{\bibfnamefont{Y.}~\bibnamefont{{Loh}}}, \bibnamefont{and}
  \bibinfo{author}{\bibfnamefont{M.~A.} \bibnamefont{{Strauss}}}
  (\bibinfo{year}{2003}).

\bibitem[{\citenamefont{{Jarrett} et~al.}(2000)\citenamefont{{Jarrett},
  {Chester}, {Cutri}, {Schneider}, {Skrutskie}, and
  {Huchra}}}]{2000AJ....119.2498J}
\bibinfo{author}{\bibfnamefont{T.~H.} \bibnamefont{{Jarrett}}},
  \bibinfo{author}{\bibfnamefont{T.}~\bibnamefont{{Chester}}},
  \bibinfo{author}{\bibfnamefont{R.}~\bibnamefont{{Cutri}}},
  \bibinfo{author}{\bibfnamefont{S.}~\bibnamefont{{Schneider}}},
  \bibinfo{author}{\bibfnamefont{M.}~\bibnamefont{{Skrutskie}}},
  \bibnamefont{and} \bibinfo{author}{\bibfnamefont{J.~P.}
  \bibnamefont{{Huchra}}}, \bibinfo{journal}{\aj}
  \textbf{\bibinfo{volume}{119}}, \bibinfo{pages}{2498} (\bibinfo{year}{2000}).

\bibitem[{\citenamefont{{Diego} et~al.}(2003)\citenamefont{{Diego}, {Silk}, and
  {Sliwa}}}]{2003MNRAS.346..940D}
\bibinfo{author}{\bibfnamefont{J.~M.} \bibnamefont{{Diego}}},
  \bibinfo{author}{\bibfnamefont{J.}~\bibnamefont{{Silk}}}, \bibnamefont{and}
  \bibinfo{author}{\bibfnamefont{W.}~\bibnamefont{{Sliwa}}},
  \bibinfo{journal}{\mnras} \textbf{\bibinfo{volume}{346}},
  \bibinfo{pages}{940} (\bibinfo{year}{2003}).

\bibitem[{\citenamefont{{Snowden} et~al.}(1997)\citenamefont{{Snowden},
  {Egger}, {Freyberg}, {McCammon}, {Plucinsky}, {Sanders}, {Schmitt},
  {Truemper}, and {Voges}}}]{1997ApJ...485..125S}
\bibinfo{author}{\bibfnamefont{S.~L.} \bibnamefont{{Snowden}}},
  \bibinfo{author}{\bibfnamefont{R.}~\bibnamefont{{Egger}}},
  \bibinfo{author}{\bibfnamefont{M.~J.} \bibnamefont{{Freyberg}}},
  \bibinfo{author}{\bibfnamefont{D.}~\bibnamefont{{McCammon}}},
  \bibinfo{author}{\bibfnamefont{P.~P.} \bibnamefont{{Plucinsky}}},
  \bibinfo{author}{\bibfnamefont{W.~T.} \bibnamefont{{Sanders}}},
  \bibinfo{author}{\bibfnamefont{J.~H.~M.~M.} \bibnamefont{{Schmitt}}},
  \bibinfo{author}{\bibfnamefont{J.}~\bibnamefont{{Truemper}}},
  \bibnamefont{and} \bibinfo{author}{\bibfnamefont{W.}~\bibnamefont{{Voges}}},
  \bibinfo{journal}{\apj} \textbf{\bibinfo{volume}{485}}, \bibinfo{pages}{125}
  (\bibinfo{year}{1997}).

\bibitem[{\citenamefont{{Fosalba} et~al.}(2003)\citenamefont{{Fosalba},
  {Gazta{\~ n}aga}, and {Castander}}}]{2003ApJ...597L..89F}
\bibinfo{author}{\bibfnamefont{P.}~\bibnamefont{{Fosalba}}},
  \bibinfo{author}{\bibfnamefont{E.}~\bibnamefont{{Gazta{\~ n}aga}}},
  \bibnamefont{and} \bibinfo{author}{\bibfnamefont{F.~J.}
  \bibnamefont{{Castander}}}, \bibinfo{journal}{\apjl}
  \textbf{\bibinfo{volume}{597}}, \bibinfo{pages}{L89} (\bibinfo{year}{2003}).

\bibitem[{\citenamefont{{Abazajian} et~al.}(2003)\citenamefont{{Abazajian},
  {Adelman-McCarthy}, {Ag{\" u}eros}, {Allam}, {Anderson}, {Annis}, {Bahcall},
  {Baldry}, {Bastian}, {Berlind} et~al.}}]{2003AJ....126.2081A}
\bibinfo{author}{\bibfnamefont{K.}~\bibnamefont{{Abazajian}}},
  \bibinfo{author}{\bibfnamefont{J.~K.} \bibnamefont{{Adelman-McCarthy}}},
  \bibinfo{author}{\bibfnamefont{M.~A.} \bibnamefont{{Ag{\" u}eros}}},
  \bibinfo{author}{\bibfnamefont{S.~S.} \bibnamefont{{Allam}}},
  \bibinfo{author}{\bibfnamefont{S.~F.} \bibnamefont{{Anderson}}},
  \bibinfo{author}{\bibfnamefont{J.}~\bibnamefont{{Annis}}},
  \bibinfo{author}{\bibfnamefont{N.~A.} \bibnamefont{{Bahcall}}},
  \bibinfo{author}{\bibfnamefont{I.~K.} \bibnamefont{{Baldry}}},
  \bibinfo{author}{\bibfnamefont{S.}~\bibnamefont{{Bastian}}},
  \bibinfo{author}{\bibfnamefont{A.}~\bibnamefont{{Berlind}}},
  \bibnamefont{et~al.}, \bibinfo{journal}{\aj} \textbf{\bibinfo{volume}{126}},
  \bibinfo{pages}{2081} (\bibinfo{year}{2003}).

\bibitem[{\citenamefont{{Hern{\' a}ndez-Monteagudo} and {Rubi{\~
  n}o-Mart{\'{\i}}n}}(2004)}]{2004MNRAS.347..403H}
\bibinfo{author}{\bibfnamefont{C.}~\bibnamefont{{Hern{\' a}ndez-Monteagudo}}}
  \bibnamefont{and} \bibinfo{author}{\bibfnamefont{J.~A.} \bibnamefont{{Rubi{\~
  n}o-Mart{\'{\i}}n}}}, \bibinfo{journal}{\mnras}
  \textbf{\bibinfo{volume}{347}}, \bibinfo{pages}{403} (\bibinfo{year}{2004}).

\bibitem[{\citenamefont{{Tegmark} et~al.}(2003)\citenamefont{{Tegmark}, {de
  Oliveira-Costa}, and {Hamilton}}}]{2003PhRvD..68l3523T}
\bibinfo{author}{\bibfnamefont{M.}~\bibnamefont{{Tegmark}}},
  \bibinfo{author}{\bibfnamefont{A.}~\bibnamefont{{de Oliveira-Costa}}},
  \bibnamefont{and} \bibinfo{author}{\bibfnamefont{A.~J.}
  \bibnamefont{{Hamilton}}}, \bibinfo{journal}{\prd}
  \textbf{\bibinfo{volume}{68}}, \bibinfo{pages}{123523}
  (\bibinfo{year}{2003}).

\bibitem[{\citenamefont{{Myers} et~al.}(2004)\citenamefont{{Myers}, {Shanks},
  {Outram}, {Frith}, and {Wolfendale}}}]{2004MNRAS.347L..67M}
\bibinfo{author}{\bibfnamefont{A.~D.} \bibnamefont{{Myers}}},
  \bibinfo{author}{\bibfnamefont{T.}~\bibnamefont{{Shanks}}},
  \bibinfo{author}{\bibfnamefont{P.~J.} \bibnamefont{{Outram}}},
  \bibinfo{author}{\bibfnamefont{W.~J.} \bibnamefont{{Frith}}},
  \bibnamefont{and} \bibinfo{author}{\bibfnamefont{A.~W.}
  \bibnamefont{{Wolfendale}}}, \bibinfo{journal}{\mnras}
  \textbf{\bibinfo{volume}{347}}, \bibinfo{pages}{L67} (\bibinfo{year}{2004}).

\bibitem[{\citenamefont{{Maddox} et~al.}(1990)\citenamefont{{Maddox},
  {Efstathiou}, {Sutherland}, and {Loveday}}}]{1990MNRAS.243..692M}
\bibinfo{author}{\bibfnamefont{S.~J.} \bibnamefont{{Maddox}}},
  \bibinfo{author}{\bibfnamefont{G.}~\bibnamefont{{Efstathiou}}},
  \bibinfo{author}{\bibfnamefont{W.~J.} \bibnamefont{{Sutherland}}},
  \bibnamefont{and}
  \bibinfo{author}{\bibfnamefont{J.}~\bibnamefont{{Loveday}}},
  \bibinfo{journal}{\mnras} \textbf{\bibinfo{volume}{243}},
  \bibinfo{pages}{692} (\bibinfo{year}{1990}).

\bibitem[{\citenamefont{{Abell} et~al.}(1989)\citenamefont{{Abell}, {Corwin},
  and {Olowin}}}]{1989ApJS...70....1A}
\bibinfo{author}{\bibfnamefont{G.~O.} \bibnamefont{{Abell}}},
  \bibinfo{author}{\bibfnamefont{H.~G.} \bibnamefont{{Corwin}}},
  \bibnamefont{and} \bibinfo{author}{\bibfnamefont{R.~P.}
  \bibnamefont{{Olowin}}}, \bibinfo{journal}{\apjs}
  \textbf{\bibinfo{volume}{70}}, \bibinfo{pages}{1} (\bibinfo{year}{1989}).

\bibitem[{\citenamefont{{Komatsu} and {Seljak}}(2002)}]{2002MNRAS.336.1256K}
\bibinfo{author}{\bibfnamefont{E.}~\bibnamefont{{Komatsu}}} \bibnamefont{and}
  \bibinfo{author}{\bibfnamefont{U.}~\bibnamefont{{Seljak}}},
  \bibinfo{journal}{\mnras} \textbf{\bibinfo{volume}{336}},
  \bibinfo{pages}{1256} (\bibinfo{year}{2002}).

\bibitem[{\citenamefont{{Hinshaw} et~al.}(2003)\citenamefont{{Hinshaw},
  {Spergel}, {Verde}, {Hill}, {Meyer}, {Barnes}, {Bennett}, {Halpern},
  {Jarosik}, {Kogut} et~al.}}]{2003ApJS..148..135H}
\bibinfo{author}{\bibfnamefont{G.}~\bibnamefont{{Hinshaw}}},
  \bibinfo{author}{\bibfnamefont{D.~N.} \bibnamefont{{Spergel}}},
  \bibinfo{author}{\bibfnamefont{L.}~\bibnamefont{{Verde}}},
  \bibinfo{author}{\bibfnamefont{R.~S.} \bibnamefont{{Hill}}},
  \bibinfo{author}{\bibfnamefont{S.~S.} \bibnamefont{{Meyer}}},
  \bibinfo{author}{\bibfnamefont{C.}~\bibnamefont{{Barnes}}},
  \bibinfo{author}{\bibfnamefont{C.~L.} \bibnamefont{{Bennett}}},
  \bibinfo{author}{\bibfnamefont{M.}~\bibnamefont{{Halpern}}},
  \bibinfo{author}{\bibfnamefont{N.}~\bibnamefont{{Jarosik}}},
  \bibinfo{author}{\bibfnamefont{A.}~\bibnamefont{{Kogut}}},
  \bibnamefont{et~al.}, \bibinfo{journal}{\apjs}
  \textbf{\bibinfo{volume}{148}}, \bibinfo{pages}{135} (\bibinfo{year}{2003}).

\bibitem[{lam()}]{lamda_cross_powspec}
\urlprefix\url{http://lambda.gsfc.nasa.gov/product/map/ang_power_spec.cfm}.

\bibitem[{\citenamefont{{Holder}}(2002)}]{2002ApJ...580...36H}
\bibinfo{author}{\bibfnamefont{G.~P.} \bibnamefont{{Holder}}},
  \bibinfo{journal}{\apj} \textbf{\bibinfo{volume}{580}}, \bibinfo{pages}{36}
  (\bibinfo{year}{2002}).

\bibitem[{\citenamefont{{Readhead} et~al.}(2004)\citenamefont{{Readhead},
  {Mason}, {Contaldi}, {Pearson}, {Bond}, {Myers}, {Padin}, {Sievers},
  {Cartwright}, {Shepherd} et~al.}}]{2004astro.ph..2359R}
\bibinfo{author}{\bibfnamefont{A.~C.~S.} \bibnamefont{{Readhead}}},
  \bibinfo{author}{\bibfnamefont{B.~S.} \bibnamefont{{Mason}}},
  \bibinfo{author}{\bibfnamefont{C.~R.} \bibnamefont{{Contaldi}}},
  \bibinfo{author}{\bibfnamefont{T.~J.} \bibnamefont{{Pearson}}},
  \bibinfo{author}{\bibfnamefont{J.~R.} \bibnamefont{{Bond}}},
  \bibinfo{author}{\bibfnamefont{S.~T.} \bibnamefont{{Myers}}},
  \bibinfo{author}{\bibfnamefont{S.}~\bibnamefont{{Padin}}},
  \bibinfo{author}{\bibfnamefont{J.~L.} \bibnamefont{{Sievers}}},
  \bibinfo{author}{\bibfnamefont{J.~K.} \bibnamefont{{Cartwright}}},
  \bibinfo{author}{\bibfnamefont{M.~C.} \bibnamefont{{Shepherd}}},
  \bibnamefont{et~al.}, \bibinfo{journal}{ArXiv e-print astro-ph/0402359}
  (\bibinfo{year}{2004}), \eprint{astro-ph/0402359}.

\bibitem[{\citenamefont{{Seljak} et~al.}(2003)\citenamefont{{Seljak},
  {McDonald}, and {Makarov}}}]{2003MNRAS.342L..79S}
\bibinfo{author}{\bibfnamefont{U.}~\bibnamefont{{Seljak}}},
  \bibinfo{author}{\bibfnamefont{P.}~\bibnamefont{{McDonald}}},
  \bibnamefont{and}
  \bibinfo{author}{\bibfnamefont{A.}~\bibnamefont{{Makarov}}},
  \bibinfo{journal}{\mnras} \textbf{\bibinfo{volume}{342}},
  \bibinfo{pages}{L79} (\bibinfo{year}{2003}).

\bibitem[{\citenamefont{{Slosar} et~al.}(2004)\citenamefont{{Slosar}, {Seljak},
  and {Makarov}}}]{2004astro.ph..3073S}
\bibinfo{author}{\bibfnamefont{A.}~\bibnamefont{{Slosar}}},
  \bibinfo{author}{\bibfnamefont{U.}~\bibnamefont{{Seljak}}}, \bibnamefont{and}
  \bibinfo{author}{\bibfnamefont{A.}~\bibnamefont{{Makarov}}},
  \bibinfo{journal}{ArXiv Astrophysics e-prints}  (\bibinfo{year}{2004}),
  \eprint{astro-ph/0403073}.

\bibitem[{\citenamefont{{Verde} et~al.}(2003)\citenamefont{{Verde}, {Peiris},
  {Spergel}, {Nolta}, {Bennett}, {Halpern}, {Hinshaw}, {Jarosik}, {Kogut},
  {Limon} et~al.}}]{2003ApJS..148..195V}
\bibinfo{author}{\bibfnamefont{L.}~\bibnamefont{{Verde}}},
  \bibinfo{author}{\bibfnamefont{H.~V.} \bibnamefont{{Peiris}}},
  \bibinfo{author}{\bibfnamefont{D.~N.} \bibnamefont{{Spergel}}},
  \bibinfo{author}{\bibfnamefont{M.~R.} \bibnamefont{{Nolta}}},
  \bibinfo{author}{\bibfnamefont{C.~L.} \bibnamefont{{Bennett}}},
  \bibinfo{author}{\bibfnamefont{M.}~\bibnamefont{{Halpern}}},
  \bibinfo{author}{\bibfnamefont{G.}~\bibnamefont{{Hinshaw}}},
  \bibinfo{author}{\bibfnamefont{N.}~\bibnamefont{{Jarosik}}},
  \bibinfo{author}{\bibfnamefont{A.}~\bibnamefont{{Kogut}}},
  \bibinfo{author}{\bibfnamefont{M.}~\bibnamefont{{Limon}}},
  \bibnamefont{et~al.}, \bibinfo{journal}{\apjs}
  \textbf{\bibinfo{volume}{148}}, \bibinfo{pages}{195} (\bibinfo{year}{2003}).

\bibitem[{\citenamefont{{Gelman} and {Rubin}}(1992)}]{gelman92}
\bibinfo{author}{\bibfnamefont{A.}~\bibnamefont{{Gelman}}} \bibnamefont{and}
  \bibinfo{author}{\bibfnamefont{D.}~\bibnamefont{{Rubin}}}
  (\bibinfo{year}{1992}).

\bibitem[{\citenamefont{{Rybicki} and {Press}}(1992)}]{1992ApJ...398..169R}
\bibinfo{author}{\bibfnamefont{G.~B.} \bibnamefont{{Rybicki}}}
  \bibnamefont{and} \bibinfo{author}{\bibfnamefont{W.~H.}
  \bibnamefont{{Press}}}, \bibinfo{journal}{\apj}
  \textbf{\bibinfo{volume}{398}}, \bibinfo{pages}{169} (\bibinfo{year}{1992}).

\end{thebibliography}

\appendix
\section{Cross-spectrum contaminant estimator} \label{sec:estderiv}
In this appendix we present a generalized version of the point source estimator
 of \citeN{2003ApJS..148..135H}, using a vector notation.  We discuss 
contaminants in general and do not mention point sources or SZ specifically.  
We assume we know the power spectrum shape and frequency dependence of all 
contaminants.  

Our data are the cross-power spectra from the experiment.  Let vector $\D = \{ 
C^\bi{i}_l \}$ be these spectra.  The multipole (or multipole bin) is denoted 
by $l$ and the cross correlation is denoted by $\bi{i} =$ W1W2, W1V1, {\em 
etc.}  
We use a Gaussian model for the likelihood $\like(\D)$ of the set of 
cross-spectra:
\be
-2 \log \like \propto \left[ \D -  \langle \D \rangle\right]\t \Sgi \left[ \D -
 \langle \D \rangle \right]. \label{eqn:like}
\ee
where the covariance $\Sg = \langle (\D-\langle \D \rangle)(\D-\langle \D 
\rangle)\t \rangle$ can be written as $\Sg = \{ \Sigma^{\bi{ii'}}_{ll'} \}$.

We postulate that the data $\D$ is the sum of the CMB and a number of 
contaminants. For each of these contributions to the data we need a model, and 
parameters for that model.  Some of these parameters we will estimate, and some
we will marginalize.

First we consider the CMB.  The parameters which describe the CMB are the band 
powers: $C_l$.  We organize these band powers into a vector $\C = \{ C_l \}$.  
We will later marginalize over these parameters.  To relate the CMB parameters 
to the data we use the window function $\beps = \{ w^\bi{i}_{ll'} \}$ for each 
cross spectrum channel pair.  We may think of the window function as the 
response of the instrument to a given set of CMB $C_l$'s.  In the absence of 
noise and contaminants, the data would be given by the matrix multiplication 
$\D = \beps\C$.  

We use as similar notation for the contaminants.  We describe each contaminant 
by a power spectrum shape, a frequency dependence, and an amplitude.  We take 
the shapes and frequency dependences as given, and the amplitudes as 
parameters.  We divide the amplitude parameters into two groups, those to 
marginalize and those to estimate.   The amplitude parameters to {\em 
marginalize} we denote by vector $\A = \{  A_\alpha \}$, where $\alpha$ runs 
over the components to marginalize.  The amplitude parameters to {\em estimate}
we denote by vector $\B = \{  B_\beta \}$, where $\beta$ runs over the 
components to estimate.  

We may think of the shape and frequency dependences as the response of the 
instrument to a given set of contaminant amplitudes.  Thus the shape dependence
already includes the influence of the window function. Similarly to the 
amplitudes, we divide the shape and frequency dependences into two groups.  The
shape and frequency dependence of the contaminants to marginalize we write as 
$\bS = \{  S^\bi{i} _{l\alpha} \}$. The shape and frequency dependence of the 
contaminants to estimate we write as $\Z = \{  Z^\bi{i} _{l\beta} \}$. In the 
absence of CMB and noise, the data would be given by $\D = \bS \A + \Z \B$.

If we include all components and uncorrelated noise, then we can write the 
expected data as 
\be
\langle \D \rangle = \beps \C + \bS \A + \Z \B.
\ee
We are considering cross-spectra only, and have included no noise term.

We substitute into the likelihood expression:
\bea\nonumber
-2 \log \like &\propto& \left[ \D - (\beps \C + \bS \A + \Z \B) \right]\t \Sgi 
\\ 
&& \qquad  \left[ \D - (\beps \C + \bS \A + \Z \B) \right]. 
\label{eqn:expandlike}
\eea
We seek to estimate $\B$.  This we accomplish by repeatedly completing the 
square in the likelihood expression, as follows.  Note that completing the 
square to marginalize out a component is equivalent to letting the covariance 
for that component tend to infinity \cite{1992ApJ...398..169R}.  Thus we 
disregard all information about that component.  The likelihood may be 
rewritten as a quadratic equation in the CMB power spectrum,
\be
-2 \log \like \propto \C\t \a \C + \frac{1}{2} (\b\t\C + \C\t\b) + \c\t\Sgi\c,
\ee
where we have introduced the shorthand
\bea \nonumber
\a &\equiv& \beps\t \Sgi \beps \\ \nonumber
\b &\equiv& -2\beps\t\Sgi( \D - \bS\A - \Z\B) \\ 
\c &\equiv& \D - \bS\A - \Z\B.
\eea
We complete the square for variable $\C$:
\bea \nonumber
-2 \log \like &\propto& \left[ \C +  \frac{1}{2} \a^{-1} \b \right] \t \a  
\left[ \C +  \frac{1}{2} \a^{-1} \b \right] \\ 
&& + \c\t\Sgi\c - \frac{1}{4} \b\t \a^{-1} \b
\eea
Note that the final term may be rewritten in terms of our auxiliary variable 
$\c$,
\be 
\frac{1}{4} \b\t \a^{-1} \b = \c\t\left(\Sgi \beps\left(\beps\t \Sgi 
\beps\right)^{-1} \beps\t \Sgi \right)\c.
\ee
Thus if we define the symmetric matrix
\be 
\E \equiv \Sgi -  \Sgi \beps\left(\beps\t \Sgi \beps\right)^{-1} \beps\t \Sgi, 
\label{eqn:E}
\ee
we may compactly express the $\log$ likelihood as a term which depends on the 
CMB $\C$, and a term which does not. 
\be
-2 \log \like \propto \left[ \C +  \frac{1}{2} \a^{-1} \b \right] \t \a  \left[
\C +  \frac{1}{2} \a^{-1} \b \right] + \c\t\E\c.
\ee
Let us define a marginalized likelihood, $\like_\C \equiv \int d\C\ \like $.  
Integrating over all $\C$, we find
\bea \nonumber
-2 \log \like_\C &\propto& \c\t\E\c \\ \nonumber
&\propto& [\D-\bS\A-\Z\B]\t \E \  \\ && \qquad [\D-\bS\A-\Z\B].
\eea
Note the similarity to equation (\ref{eqn:expandlike}).  The matrix $\E$ is the
new inverse covariance, once the CMB is marginalized out.

We wish to repeat this sequence to marginalize the variable $\A$.  Thus we 
write
\be
-2 \log \like_\C \propto \A\t\f\A +  \frac{1}{2}(\g\t\A + \A\t\g) + \h\t\E\h
\ee
where we have introduced
\bea \nonumber
\f &\equiv& \bS\t \E \bS \\ \nonumber
\g &\equiv& -2 \bS\t \E (\D - \Z\B) \\
\h &\equiv& \D - \Z\B.
\eea
Again we complete the square, now for variable $\A$:
\bea\nonumber 
-2 \log \like_\C &\propto& \left[\A +  \f^{-1} \g \right]\t \f \left[\A +  
\f^{-1} \g \right] \\ 
&&+ \h\t\E\h -  \frac{1}{4} \g\t \f^{-1} \g.
\eea
Noting the last term may be re-written in terms of $\h$,
\be
\frac{1}{4} \g\t \f^{-1} \g = \h\t\left(\E \bS\left(\bS\t \E 
\bS\right)^{-1}\bS\t \E\right)\h,
\ee
we write with analogy to equation (\ref{eqn:E}),
\be
\F \equiv \E - \E \bS\left(\bS\t \E \bS\right)^{-1}\bS\t \E. 
\label{eqn:F}
\ee
Define $\like_{\C,\A} \equiv \int  d\C d\A\ \like = \int d\A\ \like_\C $, and 
we have integrated out our contaminants:
\bea\nonumber
-2 \log \like_{\C,\A} &\propto& \h\t\F\h \\ 
&\propto& [\D - \Z\B]\t \F\ [\D - \Z\B].
\eea

To estimate the amplitudes $\B$, we express the marginalized likelihood as 
\be
-2 \log \like_{\C,\A} \propto \B\t \p \B +  \frac{1}{2}(\q\t\B + \B\t\q) + r
\ee
where we have introduced 
\bea\nonumber
\p &\equiv&  \Z\t\F\Z \\\nonumber
\q &\equiv&  - 2 \Z\t\F\D \\
r &\equiv& \D\t\F\D.
\eea
We complete the square one final time for variable $\B$:
\be
-2 \log \like_{\C,\A} \propto [ \B - \frac{1}{2} \p^{-1} \q]\t \p [ \B - 
\frac{1}{2} \p^{-1} \q] + r - \frac{1}{4} \q\t \p^{-1} \q.
\ee
Now $\like_{\C,\A}(\B) \propto \exp[-\frac{1}{2} (\B - \langle \B \rangle)\t ( 
\SgB)\in (\B - \langle \B \rangle)]$.  If we estimate
\be
\bar\B \equiv (\Z\t\F\Z)^{-1} \Z\t \F \D,
\ee
we have $\langle \bar\B \rangle = \langle \B \rangle$, which indicates our 
estimator is unbiased.  Moreover, the estimator is unbiased even if the 
original covariance $\Sg$ from equation (\ref{eqn:like}) is incorrect.  This is
shown by integrating the (flawed) estimator with the likelihood to take the 
ensemble average.  We also note the covariance on the $\B$ parameter estimates,
\be
\SgB = (\Z\t \F \Z)^{-1}.
\ee

We note that neither the estimator $\bar \B$ nor its variance $\SgB$ are 
sensitive to the cosmic variance from the CMB.  This makes sense because the 
CMB is completely projected out.  The estimator does not care about the value 
of any CMB multipole, so the cosmic variance of the multipoles is also 
immaterial.

This independence from cosmic variance may be shown algebraically.  The 
estimate appears to depend on the cosmic variance contribution to $\E$ 
(equation \ref{eqn:E}) through $\F$ (equation \ref{eqn:F}).  However, if we 
explicitly write the cosmic variance $\Sgc = \langle ( \C - \langle \C \rangle 
) ( \C - \langle \C \rangle )\t \rangle$ contribution to the variance, we see 
that it projects out of the estimate.
Let us define $\Sg^{\prime}$ as the balance of the variance, the part not due 
to cosmic variance of the CMB.  Write the covariance:
\be
\Sg = \beps \Sgc \beps\t + \Sg^{\prime},
\ee
then substitute into $\E$,
\bea\nonumber 
\E &\equiv& \left( \beps \Sgc \beps\t + \Sg^{\prime} \right)\in \\ \nonumber
&&- \left( \beps \Sgc \beps\t + \Sg^{\prime} \right)\in  \beps \\ \nonumber
&& \qquad \left(\beps\t \left( \beps \Sgc \beps\t + \Sg^{\prime} 
\right)\in \beps\right)^{-1}  \\ 
&& \qquad \qquad \beps\t \left( \beps \Sgc \beps\t + \Sg^{\prime}
\right)\in.
\eea
If we were to expand each of the matrix inverses in geometric series, we would 
find that $\E$ does not depend on $\Sgc$ at all: 
\be
\E = \left( \Sg^{\prime} \right)\in-  \left( \Sg^{\prime} \right)\in 
\beps\left(\beps\t\left( \Sg^{\prime} \right)\in \beps\right)^{-1} 
\beps\t\left( \Sg^{\prime} \right)\in .
\ee
We see that the cosmic variance of the CMB has been projected out of the 
estimator.  Thus it cannot impact the estimate $\bar \B$, or its variance 
$\SgB$.

\end{document}